\documentclass[slac_one]{revtex4}
\usepackage{graphicx}
\usepackage{fancyhdr}
\begin{document}

\title{Commissioning of the ATLAS Muon Trigger Selection} %% Paper title goes here


\author{Elisa Musto on behalf of the ATLAS Collaboration}
\affiliation{Universit\`a di Napoli ``Federico II'' and INFN Sezione di Napoli\\Via Cintia, 80126 Napoli, ITALY}
%
%\author{P. Lucas}
%\affiliation{FNAL, Batavia, IL 60510, USA}

\begin{abstract}
The performance of the three-level ATLAS muon trigger as evaluated by using LHC data is presented. Events have been selected by using only the hardware-based Level-1 trigger in order to commission and to subsequently enable the (software-based) selections of the High Level Trigger. Studies aiming at selecting prompt muons from J/$\psi$ and at reducing non prompt muon contamination have been performed. A brief overview on how the muon triggers evolve with increasing luminosity is given.
\end{abstract}

\maketitle
\thispagestyle{fancy}


\section{INTRODUCTION} % Section title should be in all capitals.
Until July 2010 the ATLAS experiment~\cite{ATLAS} \index{ATLAS} at the Large Hadron Collider (LHC)~\cite{LHC} has recorded several hundreds nb$^{ - 1}$ of proton-proton collisions at a center of mass energy ($\sqrt{s}$) of 7 TeV and a peak instantaneous luminosity of $3\cdot 10^{30}$ cm$^{ - 2}$ s$^{ - 1}$. In this early data-taking period one of the main goals of the ATLAS Collaboration has been the evaluation of the ATLAS detector~\cite{ATLAS} \index{ATLAS} performance. A wide range of interesting physics channels contain muons in the final states, so the commissioning of the Muon System (MS)~\cite{MuonSystem} \index{MuonSystem}, as well as the Muon Trigger~\cite{MuonTrigger} \index{MuonTrigger} are of primary relevance.

The ATLAS Trigger has three levels, aiming at progressively reducing the LHC collision rate of $\mathcal{O}$(GHz) to $\sim$ 200 Hz. The Level 1 (L1) Muon Trigger is hardware-based and makes use of Resistive Plate Chambers (RPC) in the barrel region ($|\eta|<1.05$) and Thin Gap Chambers (TGC) in the endcaps ($1.05<|\eta|<2.4$). It selects muon tracks coming from the Interaction Point (IP) having a transverse momentum ($p_{T}$) above a given threshold. The first estimation of the muon track parameters ($\eta$, $\phi$, $p_{T}$), named Region of Interest (RoI), is passed to the High Level Trigger (HLT), that comprises the two subsequent software-based trigger levels, Level2 (L2) and Event Filter (EF). 

At the L2 different RoIs are processed in parallel. Several algorithms are available: 
\begin{itemize}
\item $\mu$-Fast uses the trigger chambers and the precision measurements of the Muon Drift Chambers to reconstruct a muon in the MS and extrapolate it to the IP, obtaining a so-called ``Stand Alone (SA) muon''; 
\item $\mu$-Comb combines the $\mu$-Fast MS track to a reconstructed matching track in the Inner Detector (ID) in order to improve the parameters resolution: the resulting track corresponds to a ``combined muon''; 
\item $\mu$-Iso takes into account both ID and Calorimeters information to find isolated muons; 
\item $\mu$-Tile independently of the MS identifies muons by searching for minimum ionizing particle energy deposit in the Hadronic Calorimeter.
\end{itemize}
The EF uses the full event data available to select muons by means of offline reconstruction algorithms adapted to the on-line environment. Two main reconstruction strategies are available: {\it inside-out}, where an ID track is extrapolated to the MS; {\it outside-in}, where a MS track is back-extrapolated to the IP\@. In analogy to the L2, a further algorithm combines the MS and the ID tracks.

\section{COMMISSIONING OF THE ATLAS MUON TRIGGER}
Detailed studies of the ATLAS Muon Trigger commissioning are available in~\cite{TNote}. In the first part of the data-taking period the LHC interaction rate allowed to write on disk the full L1 trigger output and to commission the HLT, that was running but not rejecting events. For this purpose only data were used when the MS and the ID components were active, the solenoidal and the toroidal magnets were on, and the Muon Trigger was operational. In order to select collisions, only events containing at least three ID tracks associated with the same primary vertex having a position sufficiently close ($|z_{vertex}|<150$ mm) to the beam line were considered; moreover, since the performance was evaluated using the offline reconstruction, the muon tracks were required to be combined muons satisfying some quality criteria\footnote{$\chi^2$ $<$ 50 for the MS-ID track combination, momentum p $>$ 4 GeV, p$_T$ $>$ 2.5 GeV, a number of silicon tracker (SCT) hits greater than five and at least one hit in the pixel detector.}. The aim was to select prompt muons and to reduce the contamination of the sample arising from the on-flight decays of light mesons.

In the following some performance plots regarding the Muon Trigger performance will be shown; the efficiency ``Turn On'' curves behavior is described by the Fermi function $f= \frac{A}{1+e^{-\frac{\left(p_T - B\right)}{C}}}$, where A is the plateau value reached \\ \vspace{0.3cm}
by the curve, B is p$_T$ value at which half the plateau is reached, C is related to the rise of the curve.
 
\begin{figure}[htb]

  \begin{minipage}[b]{8cm}
    \centering    
    \includegraphics[height=3.8cm]{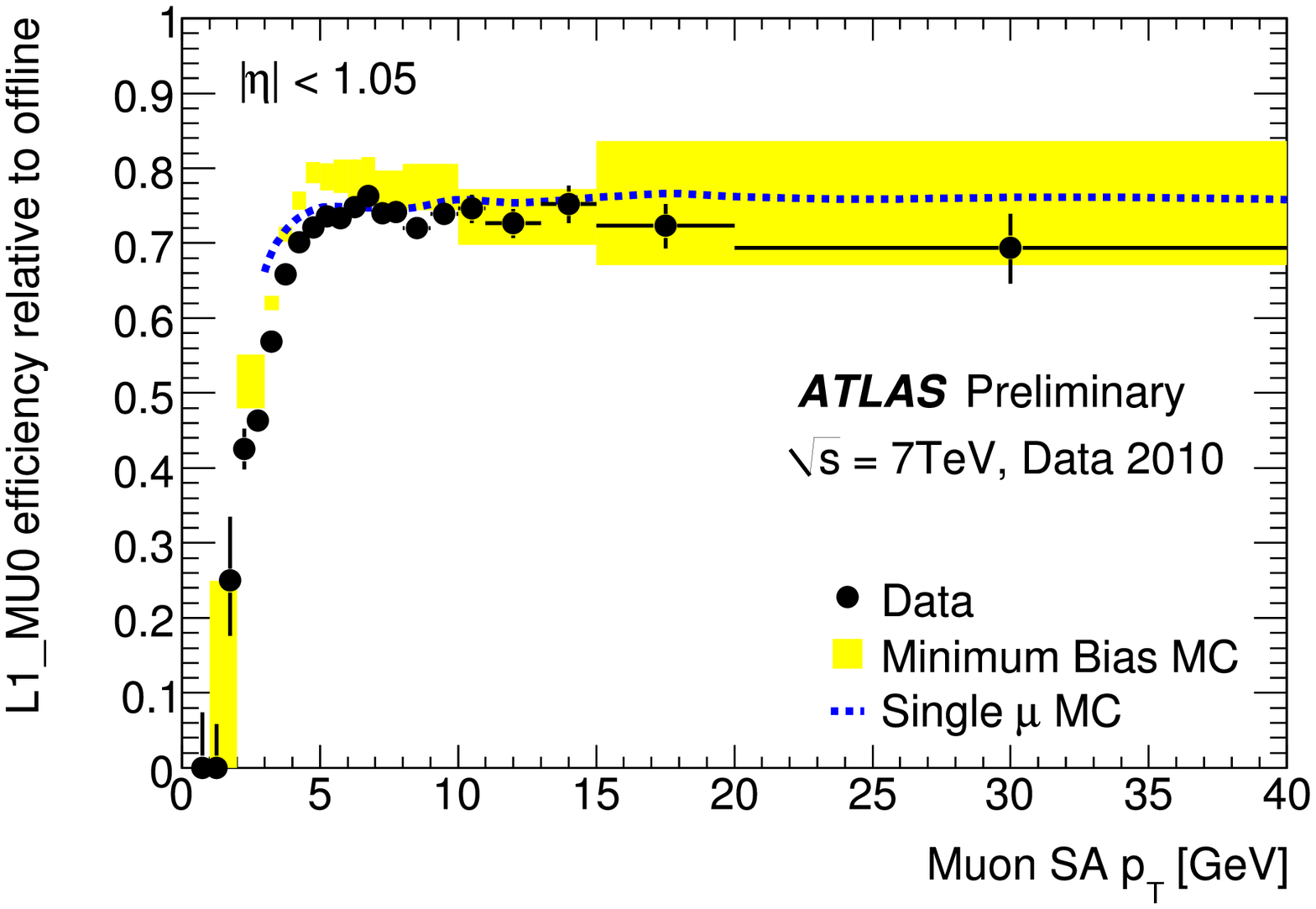}
    \caption{\small L1 RPC efficiency for the lowest threshold as a function of the SA muon p$_T$. The geometrical acceptance of the RPC trigger (80\%) explains the plateau level.\\ $\int{{\mathcal L}\cdot dt}\sim$17 nb$^{-1}$.} 
\label{fig:RPCMU0}
    \end{minipage}
   \ \hspace{2mm} \hspace{1mm}\ 
    \begin{minipage}[b]{8cm}
      \centering
      
      \includegraphics[height=3.8cm]{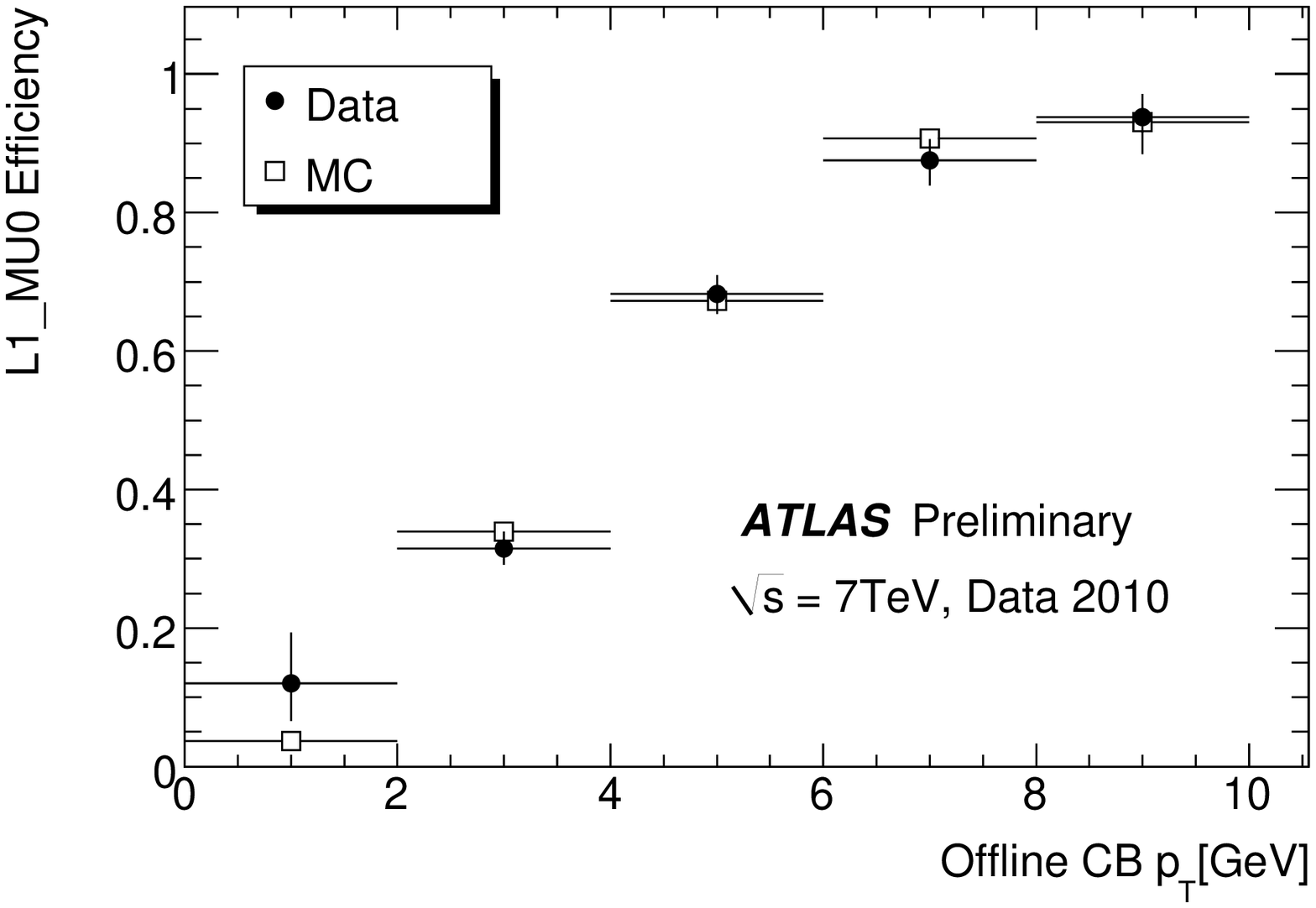}
      \caption{\small L1 Trigger efficiency for the lowest threshold as a function of the offline muon p$_T$ obtained with a ``Tag and Probe" method on J/$\psi$ candidates.\\ $\int{{\mathcal L}\cdot dt}\sim$210 nb$^{-1}$.}
      \label{fig:TP}
    \end{minipage}
  \end{figure}

Figure~\ref{fig:RPCMU0} represents the L1 muon trigger efficiency at the barrel region for the lowest (MU0) threshold compared to Monte Carlo as a function of the offline muon SA p$_T$. The angular matching criterion $\delta$R~\footnote{The angular distance between two objects is defined as: $\delta R=\sqrt{(\delta \eta)^{2} + (\delta \phi)^{2} }$} $<$ 0.5 between the offline track and the L1 RoI and the cut at 2 GeV on the SA muon p$_T$ have been applied. In Figure~\ref{fig:TP} the L1 muon trigger efficiency for the MU0 threshold evaluated with a ``Tag and Probe'' method on muons coming from J/$\psi$ decays is shown: candidates were selected by asking for two opposite sign combined muons in the invariant mass range [2860,3340] MeV being at an angular distance 0.4 $<\delta R <$ 2.0. The angular match $\delta$R $<$ 0.4 between the offline track and the L1 RoI has been used.

\begin{figure}[htb]
  \begin{minipage}[b]{8cm}
    \centering    
    \includegraphics[width=5.7cm]{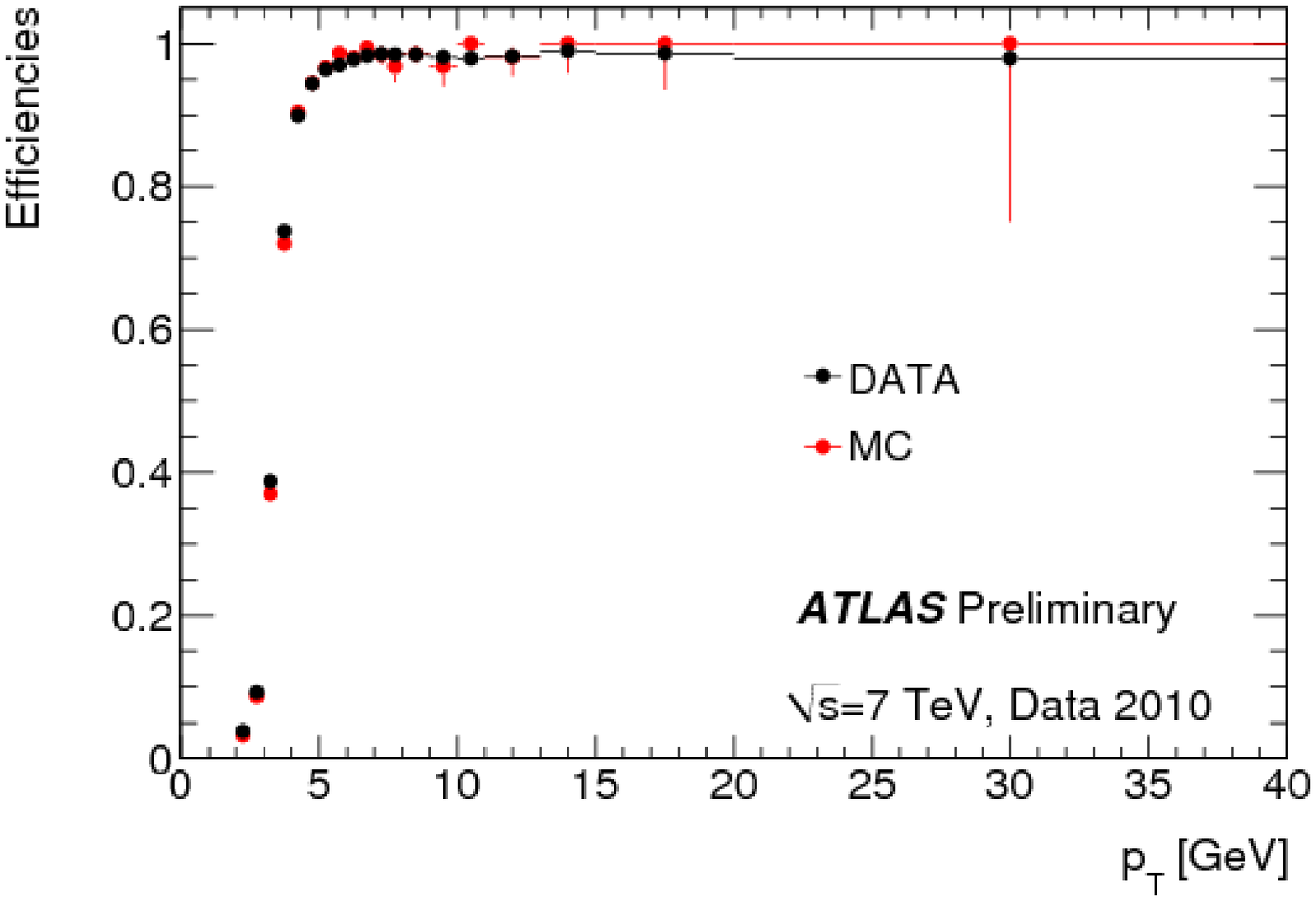}
    \caption{\small L2 $\mu$-Fast efficiency relative to the L1 trigger as a function of the offline muon p$_T$.\\ $\int{{\mathcal L}\cdot dt}\sim$94 nb$^{-1}$.} 
\label{fig:MuFast}
    \end{minipage}
   \ \hspace{2mm} \hspace{1mm}\ 
    \begin{minipage}[b]{8cm}
      \centering
      \includegraphics[width=5.7cm]{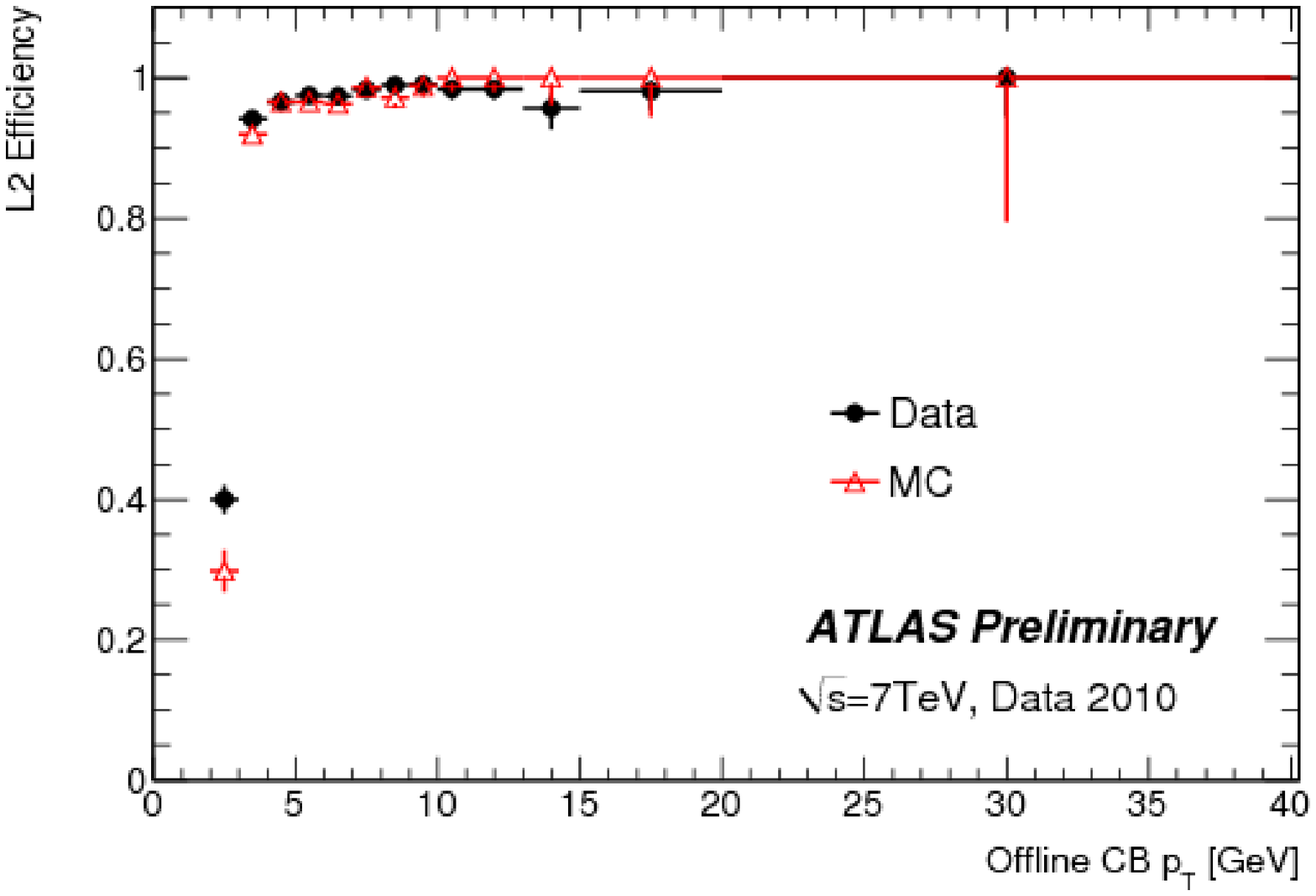}
      \caption{\small L2 $\mu$-Comb efficiency relative to the L2 $\mu$-Fast trigger as a function of the offline muon p$_T$. \\$\int{{\mathcal L}\cdot dt}\sim$94 nb$^{-1}$.}
      \label{fig:MuComb}
    \end{minipage}
  \end{figure}
\begin{figure}[htb]
  \begin{minipage}[b]{8cm}
\centering
\includegraphics[width=6 cm]{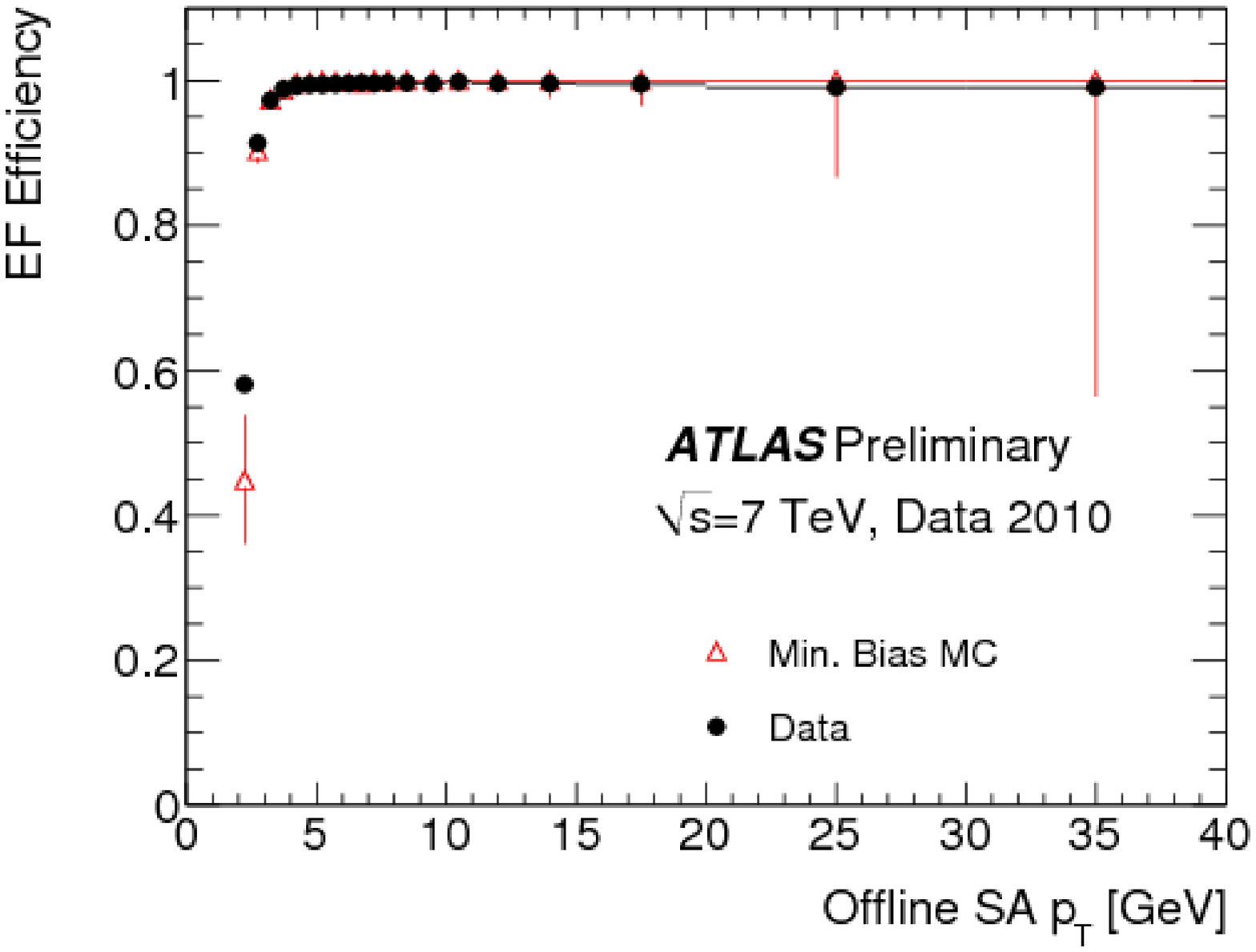}
\caption {\small EF SA efficiency relative to the offline reconstructed muons as a function of the offline SA muon p$_T$. \\$\int{{\mathcal L}\cdot dt}\sim$94 nb$^{-1}$.}
\label{fig:EF}
\end{minipage}
   \ \hspace{2mm} \hspace{1mm}\ 
  \begin{minipage}[b]{8cm}
\centering
\includegraphics[width=4.7 cm]{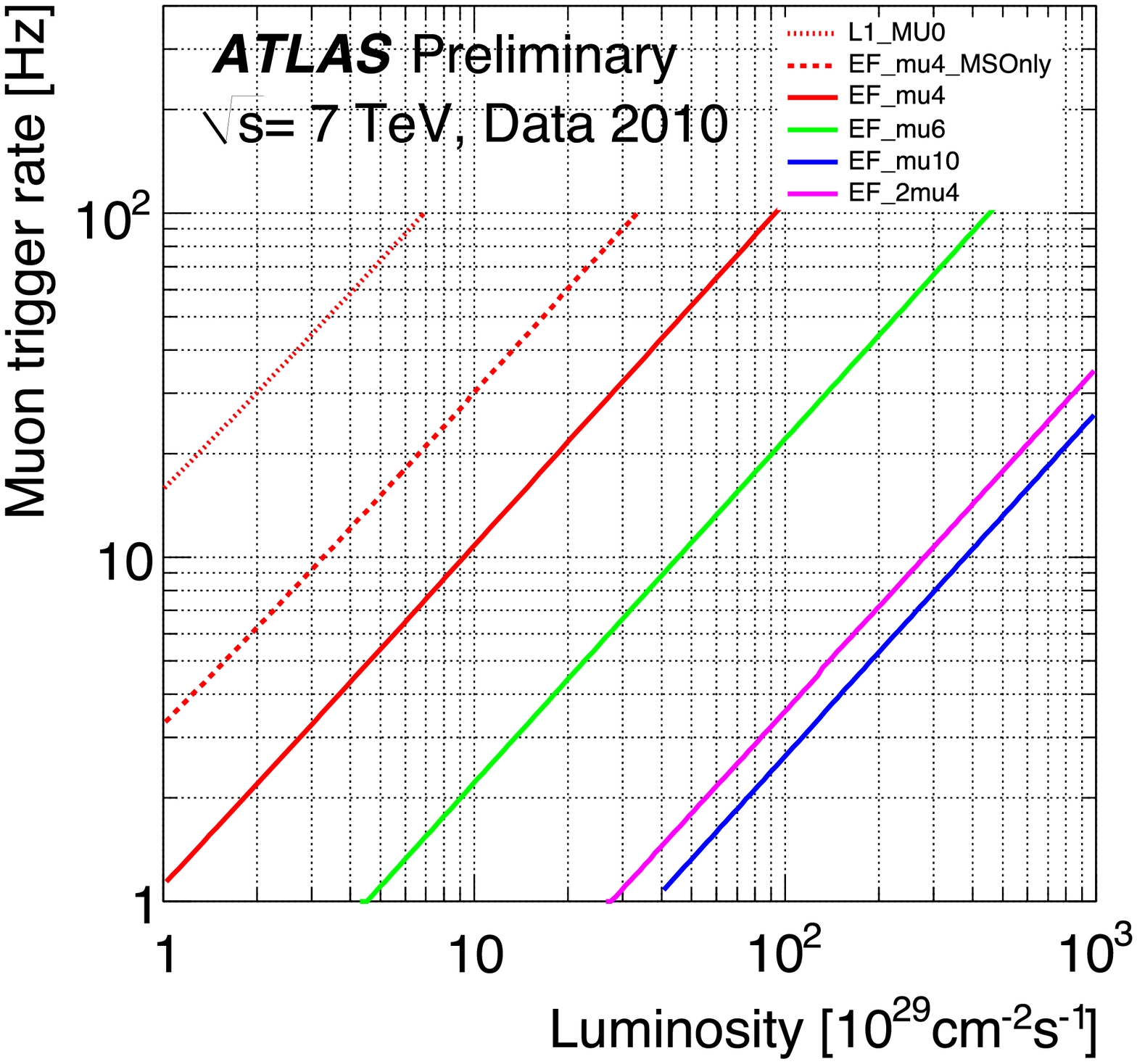}
\caption {\small Projected evolution of the trigger rates as a function of the instantaneous luminosity for different trigger thresholds.}
\label{fig:Rates}
\end{minipage}
\end{figure}

L2 and EF algorithms efficiencies were evaluated with respect to the previous trigger level: as for the L1, the results are generally in good agreement with Monte Carlo predictions. Figure~\ref{fig:MuFast} represents the L2 $\mu$-Fast efficiency evaluated with respect to the L1 trigger as a function of the offline muon $p_T$, while Figure~\ref{fig:MuComb} is the L2 $\mu$-Comb efficiency relative to L2 $\mu$-Fast as a function of the offline muon $p_T$. In both cases the selections are optimised to trigger muons with p$_{T}>$4 GeV. An example of the EF performance is reported in Figure~\ref{fig:EF}, where the efficiency of the SA selection relative to the offline reconstruction as a function of the offline SA muon $p_T$ is plotted. As for L2, thresholds are optimized to trigger muons with p$_{T}>$4 GeV. 

For what concern trigger rates (r), they depend on multiple parameters; the ones related to the collisions are: thresholds, algorithms, instantaneous luminosity ($\mathcal{L}$) and number of bunch crossings ($N_{BC}$), while cosmic rays contribute with a term proportional to $N_{BC}$. To separate these two main contributions, trigger rates can be parametrized by $r=c_{0} \mathcal{L} + c_{1}N_{BC} $, where $ c_{0}$ and $c_{1}$ are constants extracted by fits on data.
With increasing luminosity the need for reducing the trigger output rates will lead to exploit the rejection power of the HLT combined algorithms and to increase of the trigger thresholds: the evolution of the trigger rates for different trigger thresholds and levels as a function of the instantaneous luminosity is reported in Figure~\ref{fig:Rates}.

\section{CONCLUSIONS}
Data collected at LHC until July 2010 allowed to commission the three-level ATLAS Muon Trigger. Studies aiming at selecting prompt muons have been performed. Efficiencies obtained show good agreement with Monte Carlo predictions. With increasing luminosity trigger thresholds will be raised and the rejection power of the HLT will be exploited.

\end{document}